\documentclass{mem}
\usepackage{natbib}\usepackage{txfonts}\usepackage{balance}
\usepackage{graphicx}
\usepackage[a4paper]{hyperref}
\idline{75}{1}
\begin{document}
\def\teff{$T\rm_{eff }$}
\def\kms{$\mathrm {km s}^{-1}$}
\def\gtsim{\hbox{\raise 2pt \hbox {$>$} \kern-0.9em \lower 4pt \hbox {$\sim$}}}
\def\ltsim{\hbox{\raise 2pt \hbox {$<$} \kern-0.9em \lower 4pt \hbox {$\sim$}}}

\title{
Radio emission in clusters and connection to X-ray emission
}

   \subtitle{}

\author{
Luigina \,Feretti\inst{} 
          }

  \offprints{L. Feretti}

\institute{
Istituto di Radioastronomia INAF, 
Via P. Gobetti 101,
I-40129 Bologna, Italy \\
\email{lferetti@ira.inaf.it}
}

\authorrunning{Feretti }

\titlerunning{Radio emission in clusters and connection to X-ray emission}

\abstract{The most spectacular aspect of cluster radio emission is
represented by the large-scale diffuse radio sources, which cannot be
obviously associated with any individual galaxy.  These sources
demonstrate the existence of relativistic particles and magnetic
fields in the cluster volume, thus indicating the presence of
non-thermal processes in the hot intracluster medium.  The
knowledge of the properties of these sources has increased significantly
in recent years, owing to sensitive radio images and to the
development of theoretical models.  An important piece of information
on the origin and evolution of these sources can be obtained by the
cluster X-ray emission of thermal origin, and by its relation to the
radio emission.  Moreover, non-thermal X-ray emission of inverse
Compton origin gives direct information on the energy density of radio
emitting particles and the intensity of magnetic field.   } \maketitle{}

\section{Introduction}

The presence of diffuse radio emission in clusters of galaxies
demonstrates the existence of new components of the intracluster
medium (ICM), in addition to the hot gas: these are non-thermal
components, consisting of cluster-wide magnetic fields of the order of
$\sim$ 0.1-1 $\mu$G, and of a population of relativistic electrons
with Lorentz factor $\gamma >>$ 1000.  The knowledge of the properties
of non-thermal components is crucial for a comprehensive physical
description of the intracluster medium in galaxy clusters.

The diffuse radio emission shows different phenomenology depending on
the cluster evolutionary state: radio halos and relics are detected in
clusters which have recently undergone a merger event, while
mini-halos are associated with cooling core relaxed clusters. In these
clusters, moreover, powerful radio sources detected at the center
often show low brightness radio features filling cavities in the X-ray
gas. This represents a striking example of the interaction between thermal and
relativistic plasma.

The cluster dynamical activity plays an important role in the
formation and evolution of diffuse radio sources, thus the information
in the X-ray band is needed to analyze the interplay between the
thermal and non-thermal emission.  I will describe here the current
knowledge about non-thermal components of the ICM, with particular
emphasis on the information which is derived from observations in
the X-ray band.

\section{Radio sources}

\subsection{Radio halos}

\begin{figure}[]
\resizebox{6truecm}{!}{\includegraphics[clip=true,angle=-90]{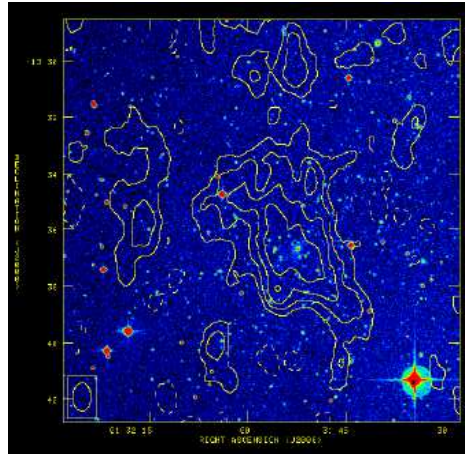}}
\caption{
\footnotesize
The giant radio halo in A209 (contours) overlayed onto the optical
map from the DSS (color). See Giovannini et al. (2006).}
\label{a209}
\end{figure}

Radio halos are diffuse radio sources of low surface brightness
permeating the central volume
of a cluster. They are typically extended, with sizes \gtsim~1 Mpc, and
are unpolarized down to a few percent level, except in A2255, where
polarized filaments have been observed at 20\%-40\% level (Govoni et al.
2005).  An example of a recently
detected  giant radio halo is in the cluster A209
(Fig. \ref{a209}, Giovannini et al. 2006). Radio halos are
typically found in clusters showing features which are indication
of  merging processes, i.e.  significant substructure, deviation
from spherical symmetry in the X-ray morphology, and strong gas
temperature gradients (Feretti 2005).  However, not all merging
clusters show giant radio halos. Radio halos 
are present in rich clusters, characterized by high X-ray
luminosities and temperatures.  Their detection rate
at the detection limit of the NRAO VLA Sky Survey (NVSS) is $ \sim$
5\% (Giovannini et al. 1999).  The detection rate increases with X-ray
luminosity to $\sim$ 35\% for clusters with X-ray luminosity larger
than 10$^{45}h_{50}^{-2}$ erg s$^{-1}$~(0.1-2.4 keV) (Giovannini \& Feretti
2002).

There are several correlation between the radio halo parameters and
parameters related to the X-ray emission of the host cluster:

i) the radio power of a halo correlates with the cluster X-ray
luminosity and consequently with the gas temperature and the total
mass (see recent results in Cassano et al. 2006). At present, it is not 
clear if X-ray bright merging clusters with no radio halo
have peculiar physical properties which do not allow the formation of a
radio halo, or if they host faint halos, undetected with the current
observational resources; 

ii) in a number of well-resolved clusters, a point-to-point
spatial correlation is observed between the radio brightness of the
halo and the X-ray brightness as detected by {\it ROSAT} (Govoni et
al. 2001, Feretti et al. 2001): a higher X-ray brightness is
associated with a higher radio brightness.  This correlation is found
also using {\it Chandra} high resolution data (Kempner \& David 2004,
Giacintucci et al. 2005).  This indicates that morphological features
detected in X-rays are strickingly similar to those found in radio,
confirming the connection between hot and relativistic plasma;

iii) the integrated radio spectra of halos are steep
%($\alpha$\footnote{S($\nu$) $\propto$ $\nu^{-\alpha}$} \gtsim~1), 
($\alpha$ \gtsim~1, with S($\nu$) $\propto$ $\nu^{-\alpha}$), 
with
a steepening at higher frequencies, as typically found in aged radio
sources. In the clusters where maps of the spectral index are
available, the radio spectrum steepens radially with the distance from
the cluster center (Giovannini et al. 1993, Feretti et al. 2004, 
Orru' et al. 2007).  In addition, it is found that the spectrum in
A665 and A2163 is flatter in the regions influenced by merger
processes (Feretti et al. 2004).  
In A2744, Orru' et al. (2007) showed that the region of highest gas
temperature is associated with the flat spectrum clump of the radio
halo, and that, in general, steep spectrum regions correlate
 with lower temperature regions.

All the above correlations favour the idea that a fraction of the
gravitational energy which is dissipated during mergers is supplied
to the halo, for the reacceleration of relativistic particles 
and amplification of
magnetic field.  Current observations are consistent with the scenario
that turbulence  following a cluster shock might be the major mechanism
responsible for the supply of energy to the electrons radiating in
radio halos.  Numerical simulations indicate that mergers can generate
strong fluid turbulence on scales of 0.1 - 1 Mpc.  The time during
which the process is effective is of $\sim$ 10$^8$ years, so that the
emission is expected to correlate with the most recent or ongoing
merger event.   Recent
theoretical developments of this aspect can be found in Blasi 2004,
Brunetti et al. (2004), Cassano \& Brunetti (2005).

\subsection{Radio relics}

\begin{figure}[]
\resizebox{\hsize}{!}{\includegraphics[clip=true,angle=-90]{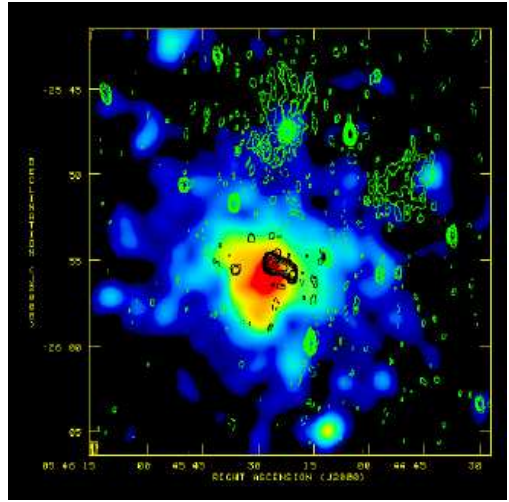}}
\caption{
\footnotesize
Image of the radio relics in A548b (contours) overlayed onto the
cluster X-ray emission from {\it ROSAT} (color). See Feretti et al. (2006).
}
\label{a548}
\end{figure}

Relic sources are diffuse extended sources similar to the radio halos
in their low surface brightness, large size (\gtsim~1 Mpc) and steep
spectrum ($\alpha$ \gtsim~ 1), but they are generally detected in the
cluster peripheral regions.  They typically show an elongated radio
structure with the major axis roughly perpendicular to the direction
of the cluster radius, and they are strongly polarized ($\sim$
20-30\%).  
A spectacular example of two likely related relics in the
same cluster is found in A548b (Fig. \ref{a548}, Feretti et al. 2006).

The detection rate of radio relics is $\sim$ 6\%, in a complete sample
of clusters (Giovannini \& Feretti 2002) at the detection limit of the
NVSS.  Relics are found in clusters both with and without a cooling
core, suggesting that they may be related to minor or off-axis
mergers, as well as to major mergers.  The radio power of relics
correlates with the cluster X-ray luminosity (Giovannini
\& Feretti 2004), as found for halos, 
although with a larger dispersion.  The existence of
this correlation indicates a link between the thermal and relativistic
plasma also in peripheral cluster regions.

Current theoretical models propose that relativistic particles
radiating in radio relics are powered by energy dissipated in shock
waves produced in the ICM during merger events.  This is consistent
with their elongated structure, almost perpendicular to the merger
axis.  This picture is supported by numerical simulations on cluster
mergers (Ricker \& Sarazin 2001, Ryu et al. 2003), which predict that
shocks forming at the cluster center at the early stages of a cluster
merger further propagate to the cluster periphery. It should be 
also mentioned the possibility that relics   
may be tracers of cosmic shock waves
related to the large-scale structure formation process, as suggested
by Bagchi et al. (2006) for A3376.

\subsection{Mini-halos}

Mini-halos are small size ($\sim$ 500 kpc) diffuse radio sources at
the center of cooling core clusters, usually surrounding a powerful
radio galaxy, as in the Perseus cluster (Sijbring 1993).  
The radio spectra of mini-halos are steep, as those of halos
and relics.  In the Perseus mini-halo, the integrated spectrum
steepens at high frequency and the spectral index distribution shows a
radial steepening (Sijbring 1993).

Gitti et al. (2002) argued that the radio emitting particles in
mini-halos cannot be connected to the central radio galaxy in terms of
particle diffusion or buoyancy, but they are likely associated with
the ICM in the cooling flow region.
The correlation observed between the mini-halo radio
power and the cooling flow power supports the idea that the mini-halos
are powered by the energy of the cooling flow (Gitti et al. 2007 and 
references therein).

A mini-halo has recently been detected in the most X-ray-luminous cluster RX
J1347.5-1145, characterized by a massive cooling flow (Gitti et al. 2007). 
This cluster follows the above correlation. In addition, it is found that 
the diffuse radio emission shows and elongation
coincident with the position of a hot sub-clump detected in X-rays.
Thus it is argued  that additional energy for the electron reacceleration
might be provided in this cluster mini-halo by the sub-merger event.

\subsection{Radio sources in X-ray cavities}

\begin{figure}[]
\resizebox{\hsize}{!}{\includegraphics[clip=true,angle=-90]{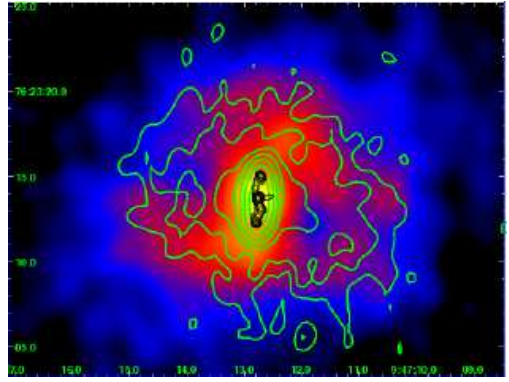}}
\caption{
\footnotesize
Radio emission in RBS797 (contours) filling X-ray cavities as detected by
{\it Chandra}. Note that the innermost radio jets are 
oriented almost perpendicularly
to the extended radio emission (Gitti et al. 2006). }
\label{rbs797}
\end{figure}

A clear example of the interaction between the radio plasma and the
hot intracluster medium was found in the {\it ROSAT} image of the Perseus
cluster (B\"{o}hringer et al. 1993), where X-ray cavities associated
with the inner radio lobes of the bright
central radio galaxy 3C84 have been first detected.  The high spatial
resolution of {\it Chandra} has confirmed the presence
of such X-ray holes (Fabian et al. 2000), coinciding with the radio
lobes and showing rims cooler than the surrounding gas.  {\it Chandra} has
permitted the detection of X-ray deficient bubbles in the inner region
of many cooling flow clusters (see e.g. Birzan et al. 2004).
A spectacular example of a radio source filling X-ray cavities is detected
in the cluster RBS797 (Fig. \ref{rbs797}, Gitti et al. 2006).

The X-ray emission in the cores of relaxed galaxy
clusters is sharply peaked and the radiative cooling time is much
shorter than the likely age of the systems, implying the
establishement of a cooling flow toward the cluster center.  However,
{\it Chandra} and {\it XMM-Newton} spectra have shown that the
effective cooling rate is greatly reduced, thus some heating process
must be taking place.  The AGN in the central galaxy is the most
viable mechanism for heating the core regions of clusters.  Indeed,
most of the dominant galaxies have active radio sources, which are
obviously blowing bubbles of relativistic plasma in the central X-ray
gas, mostly in the form of mechanical energy in jets.

To investigate the energy exchange between the central galaxy and the
cooling gas, a detailed radio - X-ray study is needed.  Moreover, the
radio bubbles are magnetized. It is expected that when they get mixed
with the IGM, they are likely to contribute to the intracluster
magnetic field.

\section {Cluster magnetic fields}

The existence of ICM magnetic fields, directly demonstrated by the
diffuse synchrotron radio emission, can also be proved by studies
of the Faraday rotation measure (RM) of polarized radio sources both
within and behind clusters. Since the RM is 
related to the cluster electron density $n_{\rm e}$,
and to the magnetic field $B_{\parallel}$ along the line of sight
$l$ , through the relation: 
${\rm RM} \propto \int\limits_0^L n_{\rm e}  B_{\parallel} dl$,
the interpretation of RM data, and consequently the inference of the
magnetic field strenght, relies on the determination of the density of
the ionized medium along the line of sight, which is obtained from X-ray data. 

Overall, the data are consistent with cluster atmospheres containing
magnetic fields in the range of 1-5 $\mu$G, regardless of the presence
or not of diffuse radio emission (see Govoni \& Feretti 2004 for a review). 
At the center of cooling core
clusters, magnetic field strenghts can be much larger, up to a few
tens $\mu$G. Magnetic fields, however, are not constant and uniform through
the cluster, thus  the 
magnetic field structure, coherence lenght, radial decline, relation 
between magnetic field strenght and gas density must be investigated.

Detailed and sensitive RM data, which will be obtained with future
generation radio telescopes, will be of little use without the X-ray
information.  Future X-ray maps of the X-ray sky at low energies will
provide a precise knowledge of the X-ray surface brightness of
clusters, i.e. of their thermal gas density, allowing the accurate and
correct interpretation of the sensitive RM measurements.

\section {Non-thermal X-ray emission}

Non-thermal phenomena can be directly studied in the X-ray band,
through the detection of X-ray inverse Compton emission due to the
scattering between the radio-emitting electrons and the CMB
photons. This emission falls in the hard X-ray domain, owing to the
very high energy of radio emitting electrons ($\gamma$
$\sim$ 10$^4$).  Since the X-ray and radio emissions are produced by
the same population of electrons undergoing inverse Compton and
synchrotron energy losses, respectively, the ratio between the X-ray
and the radio luminosities is proportional to the ratio between the
CMB and the magnetic field energy densities.  Thus the comparison
between radio and hard X-ray emission enables the determination of the
electron density and of the mean magnetic field directly, without
invoking the equipartition assumptions needed for the determination of 
these parameters from radio data.

A significant breakthrough in the measurement of hard X-ray emission
was obtained owing to the improved sensitivity and wide
spectral capabilities of the BeppoSAX and the Rossi X-ray Timing
Explorer (RXTE) satellites (see e.g. Fusco-Femiano et al. 2003, 
Govoni \& Feretti 2004, and
references therein). However, the data still refer to a handful of
clusters.  Measurements on large samples of objects will be crucial.
In addition, images of the hard X-ray distribution will be
particularly valuable to understand the origin of radiating electrons
and check reacceleration models. Indeed, under the assumption of
electron reacceleration, Brunetti et al. (2001) argue that most of the X-ray
non-thermal emission in the Coma cluster 
is produced in the outer volume,
i.e. between 30-50 arcmin from the cluster center. This is the region
which contains the large majority of the relativistic electrons able
to scatter the cosmic microwave background photons, and where the
magnetic field strength is lower than at the center.

\section {Future prospects}

Diffuse radio emission demonstrating the existence of relativistic
particles in the ICM is detected in both merging and relaxed clusters,
although with different phenomenology.  At the detection limits of the
present radio telescopes, not all merging clusters show halos or
relics, and not all cooling core clusters show mini-halos at their
centers.  Thus it is possible that there are two classes of clusters,
those hosting relativistic particles and those without relativistic
particles. Alternatively, relativistic particles may be quite common
in the cluster volume, but more sensitivity is needed to detect their
radio emission.

The study of the non-thermal processes in clusters is basically
carried out at radio waveleghts, however the understanding of these
phenomena strictly relies on the information at other wavelenths, from
the optical to gamma-ray, with a large impact coming from the
X-ray band.

Future prospects to shed light on this field in the radio regime
with next generation radio telescope (LOFAR, LWA, SKA)
include: i) search for new sources of the different classes, in
particular at low powers and at large redshifts, to improve
the information on the statistical properties of cluster diffuse
sources;  ii) polarization information on halos to get direct
information on the magnetic field structure and degree of ordering;
iii) accurate integrated spectra on a large frequency range and
detailed spectral index distributions at high resolution; iv)
detailed knowledge of the magnetic field strength and structure,
through rotation measure studies of embedded and background radio
sources.

At the same time, observations in the X-ray will be crucial to: i)
establish the cluster conditions, the merger evolutionary stage, the
presence and properties of shocks, the signatures of cluster
turbulence, and compare with radio structures and spectra; 
ii) get information on the faint peripheral cluster regions where
relics are located; iii) compare
radial profiles of the radio surface brightness and the X-ray
brightness; iv) analyse correlations between radio power and cluster
parameters (mass, X-ray luminosity, cooling flow power, etc.)  on
large samples and over a large range of parameters and at different
redshifts, to compare with expectations from theoretical models; v)
get accurate determinations of the X-ray gas density, for a correct
inference of magnetic field strength and structure from RM data; v)
obtain measurements and images of the non-thermal inverse Compton
emission in the hard X-ray domain (particularly relevant here is SIMBOL-X)
on several
clusters, with and without diffuse emission, to get independent
information on the existence of relativistic particles, and their
location.

\begin{acknowledgements}
I am grateful to the organizers for the invitation to this interesting
and fruitful meeting.
\end{acknowledgements}

\bibliographystyle{aa}

\begin{thebibliography}{}

\bibitem[{authors et al. (2000)}]{authors}
Bagchi, J., Durret, F., Lima Neto, G.B., Surajit, P.,
2006, Science, 314, 791

\bibitem[{authors et al. (2000)}]{authors}
Birzan, L., Rafferty, D.A., McNamara, B.R., Wise, M.W., Nulsen, P.E.J.,
2004, ApJ, 607, 800

\bibitem[{authors et al. (2000)}]{authors}
Blasi P., Jour. Kor. Astr. Soc., 2004,  37, 483 

\bibitem[{authors et al. (2000)}]{authors}
B\"ohringer, H., Voges, W., Fabian, A.C., Edge, A.C., Neumann, D.M.,
1993, MNRAS, 264, L25

\bibitem[{authors et al. (2000)}]{authors}
Brunetti, G., Setti, G., Feretti, L., Giovannini, G., 2001,
MNRAS, 320, 365

\bibitem[{authors et al. (2000)}]{authors}
Brunetti G., Blasi P., Cassano R., Gabici S., 2004, 
MNRAS, 350,  1174 

\bibitem[{authors et al. (2000)}]{authors}
Cassano R., Brunetti G., 2005, MNRAS 357, 1313

\bibitem[{authors et al. (2000)}]{authors}
Cassano R., Brunetti G., Setti G.,  2006, MNRAS, 369, 1577

\bibitem[{authors et al. (2000)}]{authors}
Fabian, A.C., Sanders, J.S., Ettori, S., 
%Taylor, G.B., Allen, S.W., Crawford, C.S., Iwasawa, K., 
%Johnstone, R.M., Ogle, P.M., 
et al., 2000, MNRAS, 318, L65

\bibitem[{authors et al. (2000)}]{authors}
Feretti L., Fusco-Femiano R., Giovannini G., Govoni F., 2001,
A\&A, 373, 106


\bibitem[{authors et al. (2000)}]{authors}
Feretti L., Orr\'u E., Brunetti G., Giovannini G.,
Kassim N., G. Setti G., 2004,
A\&A, 423, 111

\bibitem[{authors et al. (2000)}]{authors}
Feretti, L., 2005, AdSpR, 36, 729

\bibitem[{authors et al. (2000)}]{authors}
Feretti, L., Bacchi, M., Slee, O.B., Giovannini, G., Govoni, F., 
Andernach, H., Tsarevsky, G., 2006, MNRAS, 368, 544

\bibitem[{authors et al. (2000)}]{authors}
Fusco-Femiano, R., Dal Fiume, D., Orlandini, M., et al., 2003,
in: {\it Matter and Energy
in Clusters of Galaxies}, Eds. S. Bowyer \& C.Y. Hwang,
ASP Conference Series  Vol. 301,  p. 109

\bibitem[{authors et al. (2000)}]{authors}
Giacintucci, S., Venturi, T., Brunetti, G., 
%Bardelli, S., Dallacasa, D., 
%Ettori, S., Finoguenov, A., Rao, A.P., Zucca, E., 
et al., 2005, A\&A, 440, 867

\bibitem[{authors et al. (2000)}]{authors}
Giovannini G., Feretti L., Venturi T., Kim K.-T., Kronberg P.P,
1993,  ApJ, 406, 399

\bibitem[{authors et al. (2000)}]{authors}
Giovannini, G., Tordi, M., Feretti, L., 1999, 
New Astron.,  4, 141 

\bibitem[{authors et al. (2000)}]{authors}
Giovannini G., Feretti L., 2002,
in: {\it Merging Processes of Galaxy
Clusters}, eds. L. Feretti, I.M. Gioia \& G. Giovannini, ASSL,
Kluwer Ac. Publish., p. 197

\bibitem[{authors et al. (2000)}]{authors}
Giovannini G., Feretti L., Jour. Kor. Astr. Soc., 2004, 37, 323 

\bibitem[{authors et al. (2000)}]{authors}
Giovannini, G., Feretti, L., Govoni, F., Murgia, M., Pizzo, R., 
2006, AN, 327, 563

\bibitem[{authors et al. (2000)}]{authors}
Gitti M., Brunetti G., Setti G., 2002,
A\&A 386, 456

\bibitem[{authors et al. (2000)}]{authors}
Gitti, M., Feretti, L., Schindler, S., 2006, A\&A, 448, 853

\bibitem[{authors et al. (2000)}]{authors}
Gitti, M., Ferrari, C., Domainko, W., Feretti, L., Schindler, S., 2007,
A\&A, 470, L25

\bibitem[{authors et al. (2000)}]{authors}
Govoni, F., En{\ss}lin, T.A.,  Feretti, L., Giovannini, G., 2001, 
A\&A,  369, 441

\bibitem[{authors et al. (2000)}]{authors}
Govoni, F., Feretti, L., 2004,
Int. J. Mod. Phys. D, Vol., 13, N. 8, p. 1549

\bibitem[{authors et al. (2000)}]{authors}
Govoni, F., Murgia, M., Feretti, L., Giovannini, G., Dallacasa, D., 
Taylor, G.B., 2005, A\&A, 430, L5

\bibitem[{authors et al. (2000)}]{authors}
Kempner J.C., David L.P., 2004, MNRAS, 349, 385

\bibitem[{authors et al. (2000)}]{authors}
Orr\'u, E., Murgia, M., Feretti, L., Govoni, F., Brunetti, G., Giovannini, G., 
Girardi, M., Setti, G., 2007, A\&A, 467, 943

\bibitem[{authors et al. (2000)}]{authors}
Ricker P.M., Sarazin C.L., 2001, ApJ 561, 621 

\bibitem[{authors et al. (2000)}]{authors}
Ryu D., Kang H., Hallman E., Jones T.W.,  2003, ApJ, 593, 599 

\bibitem[{authors et al. (2000)}]{authors}
Sijbring L.G., 1993, {\it A radio continuum and HI line study of the
Perseus cluster}, PhD Thesis, University of Groningen

\end{thebibliography}

\end{document}